\documentclass[12pt]{article} \usepackage{amssymb,latexsym}
\usepackage{amsmath}
 
\columnwidth\textwidth

\begin{document}

\newtheorem{df}{Definition} \newtheorem{thm}{Theorem} \newtheorem{lem}{Lemma}
\newtheorem{rl}{Rule}
\begin{titlepage}
 
\noindent
 
\begin{center} {\LARGE Quantum measurement problem and cluster separability}
\vspace{1cm}

P. H\'{a}j\'{\i}\v{c}ek \\ Institute for Theoretical Physics \\ University of
Bern \\ Sidlerstrasse 5, CH-3012 Bern, Switzerland \\ hajicek@itp.unibe.ch

\vspace{1cm}
 
 
July 2010 \\ \vspace{1cm}
 
 
\vspace*{2cm}
 
\nopagebreak[4]
 
\begin{abstract} A modified Beltrametti-Cassinelli-Lahti model of measurement
apparatus that satisfies both the probability reproducibility condition and
the objectification requirement is constructed. Only measurements on
microsystems are considered. The cluster separability forms a basis for the
first working hypothesis: the current version of quantum mechanics leaves open
what happens to systems when they change their separation status. New rules
that close this gap can therefore be added without disturbing the logic of
quantum mechanics. The second working hypothesis is that registration
apparatuses for microsystems must contain detectors and that their readings
are signals from detectors. This implies that separation status of a
microsystem changes during both preparation and registration. A new rule that
specifies what happens at these changes and that guarantees the
objectification is formulated and discussed. A part of our result has certain
similarity with 'collapse of the wave function'.
\end{abstract}

\end{center}

\end{titlepage}

\section{Introduction} Discussions about the nature of quantum measurement
were started already by founding fathers of the theory, persisted throughout
and seem even to amplify at the present time.

An old approach to the problem of quantum measurement is Bohr's (its newer,
rigorously reformulated version is Ref.\ \cite{ludwig1}). This approach denies
that measuring apparatuses, and all classical systems in general, are quantum
systems in the sense that all their properties can be derived from, or are
compatible with, quantum mechanics. They must be described by other theories,
called pretheories. Of course some classical properties of macroscopic systems
can be obtained by quantum statistics. Ref.\ \cite{ludwig2} show that such
occasional applications of quantum mechanics to classical systems are
compatible with the form of denying the universality of quantum mechanics
specified there.

Modern approaches assume the universality of quantum mechanics together with
various further ideas. An example is the quantum decoherence theory
\cite{zeh,zurek}, another the superselection sectors approach
\cite{hepp,primas}, etc. However, the problem is far from being satisfactorily
solved by any of the modern theories. Analysis of Refs.\
\cite{d'Espagnat,bub,BLM}, as well as of our previous papers
\cite{PHJT,hajicek}, give an account of their shortcomings. In the present
paper, we adopt the definition of the problem and the proof that it is far
from being solved from Ref.\ \cite{BLM}.

Our starting point is the realist interpretation of quantum mechanics of Ref.\
\cite{PHJT} as well as the quantum theory of classical systems of Refs.\
\cite{PHJT,hajicek}. To solve the quantum measurement problem, additional
ideas seem necessary and we propose some such ideas in the present paper. They
might work in general, but we consider here only a special case. First, we
assume the validity of non-relativistic quantum mechanics. Second, we restrict
ourselves to measurements performed on microsystems such as elementary
particles or systems composite of few particles. There are other systems on
which recently a lot of interesting experiments have been done, such as
Bose-Einstein condensates, strong laser beams or currents in superconductor
rings. Such quantum states of 'large' systems, sometimes even macroscopic,
will be ignored here. Third, we shall work within a simplified theoretical
model of measurement due to Beltrametti, Cassinelli and Lahti
\cite{belt}. Fourth, our theory will consider only those registrations in
which the reading of registration apparatus is an electronic signal from a
detector.

The main idea of the paper is a new assessment of the role that the existence
of indistinguishable microsystems plays in general methods of quantum
mechanics and in the special case of preparation and registration processes.
Quantum systems can be divided into two classes according to the method of
their description. First, there are particles and systems composite of
particles of different type. Any of these and of their subsystems is a
full-fledged quantum system ${\mathcal S}$ possessing a Hilbert space
${\mathcal H}_{\mathcal S}$.  ${\mathcal H}_{\mathcal S}$ determines set of
states ${\mathcal T}({\mathcal H}_{\mathcal S})^+_1$ (positive operators with
trace 1) and set of effects ${\mathcal L}({\mathcal H}_{\mathcal S})^+_{\leq
1}$ (positive operators with norm bounded by 1 such as projections) from which
its observables are constructed (for details, see \cite{ludwig1,BLM}).  The
existence of this description and its physical meaning enabled us in Ref.\
\cite{PHJT} to view ${\mathcal S}$ as a physical object. Indeed, ${\mathcal
S}$ has a sufficient number of objective properties because e.g.\ any element
of ${\mathcal T}({\mathcal H}_{\mathcal S})^+_1$ can serve as a prepared state
of ${\mathcal S}$, and is then an objective property of ${\mathcal S}$ (for
discussion, see Refs.\ \cite{PHJT,hajicek}). However, there are also systems
composite of more than one particle of the same type. Then, there is only a
common formal one-particle Hilbert space from which a physical Hilbert space,
states and observables of the whole system are constructed. Only the
observables of the whole system are measurable. Thus, while the whole system
is an object, none of the particles is. They are described in a different way,
we call them {\em subobjects} and they form the second class of quantum
systems.

Only few textbooks (such as Ref.\ \cite{peres}) mention that these two modes
of description contain a germ of contradiction (even without realist
interpretations). Indeed, if we realise that the world is composite of many
particles and that particles of each type occur in a huge number, then the
justification of description of any such particle as an object, i.e., as if
there were no other particles of the same type, seems to be strange.
Nevertheless, such description can be justified and one justification is based
on the idea of cluster separability of Ref.\ \cite{peres}, P. 128. We
reformulate this idea, introduce the notion of separation\footnote{To prevent
misunderstanding, let us mention that the term 'nonseparability' is sometimes
used in a completely different sense (e.g., Ref.\ \cite{d'Espagnat}, P. 131)
expressing the following valid property of quantum mechanics: a quantum state
of a composite system contains more information than the sum of informations
in the states of its constituents does.} status, such as that of object or
subobject, and find that there are consequences which can have some bearing on
the quantum measurement problem.

If one applies the rules of ordinary quantum mechanics to microsystems that
change their separation status, one can obtain wrong results. The theory
cannot be expected to give reliable predictions in these cases. Our strategy
in dealing with this problem will be first to calculate as if the ordinary
quantum mechanics were applicable and then to see whether the observational
evidence suggests any corrections. From the formally logical point of view,
the current version of quantum mechanics ought to be understood as a theory of
systems that have a fixed separation status and is thus incomplete. Hence,
there is a possibility to add new rules to it without interference with its
own notions and rules.

The plan of the paper is as follows. Sec.\ 2 summarises the
Beltrametti-Cassinelli-Lahti model, defines the quantum-measurement problem
and sketches a simple no-go theorem, using ideas of Ref.\ \cite{BLM}. Sec.\ 3
analyses experiments with microsystems in order to motivate the assumption
that reading of any real registration apparatus is a signal from a
detector. This makes registration processes nearer to practice and, more
importantly, it allows us to show that a microsystem changes its separation
status during registration.

Sec.\ 4 reformulates the idea of cluster separability of Ref.\ \cite{peres} in
more rigorous terms. This facilitates the introduction of the key notions of
the paper: the separation status of a microsystem and its changes. In Sec.\ 5,
Beltrametti-Cassinelli-Lahti model is modified so that it can describe a
simplified ideal detector and corrected by adding a new rule, Rule 2. It is
based on empirical observations, in particular on the well-known fact that any
individual registration yields a definite value (the objectification
requirement). In the formulation of Rule 2, correlations play an important
role. Appendix A describes the mathematical construction of $D$-local
observables and Appendix B contains a proof that an entangled vector state of
a composite system is completely determined by correlations between
observables of a certain set.

The proposed Rule 2 is rather special and it is clear that a more general
rule, or more rules, will be necessary to make quantum mechanics
complete. This will require further work, both theoretical and
experimental. This and other questions are discussed in the Conclusion.

\section{ Beltrametti-Cassinelli-Lahti model of quantum measurement} In this
section, we are going to recapitulate the well-known ideas on measurement that
will be needed later. A summary is \cite{BLM}, P. 25: \begin{quote} \dots the
object system ${\mathcal S}$, prepared in a state ${\mathsf T}$ is brought
into a suitable contact---a {\em measurement coupling}---with another,
independently prepared system, the {\em measurement apparatus} from which the
{\em result} related to the measured observable ${\mathsf O}$ is {\em
determined} by {\em reading} the value of the {\em pointer observable}.
\end{quote} In Ref.\ \cite{BLM}, these ideas are developed in detail with the
help of models. One of them is as follows (P. 38). Let a discrete observable
${\mathsf O}$ of system ${\mathcal S}$ with Hilbert space ${\mathcal
H}_{\mathcal S}$ be measured. Let $o_k$ be eigenvalues and $\{\phi_{kj}\}$ be
the complete orthonormal set of eigenvectors,
$$
{\mathsf O}\phi_{kj} = o_k \phi_{kj}
$$
of ${\mathsf O}$. The projection ${\mathsf E}^{\mathsf O}_k$ on the eigenspace
of $o_k$ is then ${\mathsf E}^{\mathsf O}_k = \sum_j
|\phi_{kj}\rangle\langle\phi_{kj}|$. Let the registration
apparatus\footnote{In our language, a measurement consists of preparation and
registration so that what Ref.\ \cite{BLM} often calls 'measurement' is our
'registration'.} be a quantum system ${\mathcal A}$ with Hilbert space
${\mathcal H}_{\mathcal A}$ and an observable ${\mathsf A}$. Let ${\mathsf A}$
be a non-degenerate, discrete observable with the same eigenvalues $o_k$ and
with the complete orthonormal set of eigenvectors $\psi_k$,
$$
{\mathsf A}\psi_k = o_k \psi_k\ .
$$
The projection on an eigenspace is ${\mathsf E}^{\mathsf A}_k =
|\psi_k\rangle\langle\psi_k|$. ${\mathsf A}$ will be the {\em pointer
observable}.

Let the measurement start with the preparation of ${\mathcal S}$ in state
${\mathsf T}$ and the independent preparation of ${\mathcal A}$ in state
${\mathsf T}_{\mathcal A}$. The initial state of the composed system
${\mathcal S} + {\mathcal A}$ is thus ${\mathsf T}\otimes {\mathsf
T}_{\mathcal A}$.

Let ${\mathcal S}$ and ${\mathcal A}$ then interact for a finite time by the
so-called {\em measurement coupling} and let the resulting state be given by
${\mathsf U}({\mathsf T}\otimes {\mathsf T}_{\mathcal A}){\mathsf U}^\dagger$,
where ${\mathsf U}$ is a unitary transformation on ${\mathcal H}_{\mathcal
S}\otimes {\mathcal H}_{\mathcal A}$.

The final state of the apparatus is $tr_{\mathcal S}\bigl[{\mathsf U}({\mathsf
T}\otimes {\mathsf T}_{\mathcal A}){\mathsf U}^\dagger\bigr]$, where
$tr_{\mathcal S}$ is the partial trace over states of ${\mathcal S}$. The
first requirement on the model is that this state gives the same probability
measure for the pointer observable as the initial state ${\mathsf T}$
predicted for the observable ${\mathsf O}$:
$$
tr[{\mathsf T}{\mathsf E}^{\mathsf O}_k] = tr\bigl[tr_{\mathcal S}[{\mathsf
U}({\mathsf T}\otimes {\mathsf T}_{\mathcal A}){\mathsf U}^\dagger]{\mathsf
E}^{\mathsf A}_k\bigr]\ .
$$
This is called {\em probability reproducibility condition}. Now, there is a
theorem \cite{belt}:
\begin{thm} Let a measurement fulfil all assumptions and conditions listed
above. Then, for any initial vector state $\psi$ of ${\mathcal A}$, there is a
set $\{\varphi_{kl}\}$ of unit vectors in ${\mathcal H}_{\mathcal S}$
satisfying the orthogonality conditions
$$
\langle \varphi_{kl}|\varphi_{kj}\rangle = \delta_{lj}
$$
such that ${\mathsf U}$ is a unitary extension of the map
\begin{equation}\label{unitar} \phi_{kl}\otimes \psi \mapsto
\varphi_{kl}\otimes \psi_k\ .
\end{equation}
\end{thm}

One assumes further that the eigenvalues of the pointer observable are
uniquely associated with what will be read on the apparatus after the
measurement. Then, the second requirement on the model is that it has to lead
to a definite result. More precisely, the apparatus must be in one of the
states $|\psi_k\rangle\langle\psi_k|$ after each individual registration. This
is called {\em objectification requirement}. Ref.\ \cite{BLM} introduces a
more general concept of measurement that leaves open whether the
objectification requirement is satisfied or not. Such a procedure is called
{\em premeasurement}. A measurement is then a premeasurement that satisfies
objectification requirement.

Suppose that the initial state of ${\mathcal S}$ is an eigenstate, ${\mathsf
T} =|\phi_{kl}\rangle\langle\phi_{kl}|$, with the eigenvalue $o_k$. Then, Eq.\
(\ref{unitar}) implies that the final state of apparatus ${\mathcal A}$ is
$|\psi_k\rangle\langle\psi_k|$, and the premeasurement does lead to a definite
result. However, suppose next that the initial state is an arbitrary vector
state, ${\mathsf T} =|\phi\rangle\langle\phi|$. Decomposing $\phi$ into the
eigenstates,
$$
\phi = \sum_{kl} c_{kl}\phi_{kl}\ ,
$$
we obtain from Eq.\ (\ref{unitar})
\begin{equation}\label{finalSA} {\mathsf U} (\phi \otimes \psi) = \sum_k
\sqrt{p^{\mathsf O}_\phi(o_k)}\Phi_k\otimes \psi_k\ ,
\end{equation} where
\begin{equation}\label{Phik} \Phi_k = \frac{\sum_l
c_{kl}\varphi_{kl}}{\sqrt{\langle \sum_l c_{kl}\varphi_{kl}|\sum_j
c_{kj}\varphi_{kj}\rangle}}
\end{equation} and
$$
p^{\mathsf O}_\phi(o_k) = \left\langle \sum_l c_{kl}\varphi_{kl}\Biggm|\sum_j
c_{kj}\varphi_{kj}\right\rangle
$$
is the probability that a registration of ${\mathsf O}$ performed on vector
state $\phi$ gives the value $o_k$. The final state of apparatus ${\mathcal
A}$ then is
\begin{equation}\label{finalA} tr_{\mathcal S}[{\mathsf U}({\mathsf T}\otimes
{\mathsf T}_{\mathcal A}){\mathsf U}^\dagger] = \sum_{kl} \sqrt{p^{\mathsf
O}_\phi(o_k)}\sqrt{p^{\mathsf O}_\phi(o_l)}\langle\Phi_k|\Phi_l\rangle
|\psi_k\rangle\langle\psi_l|\ .
\end{equation}

Because of the orthonormality of $|\psi_k\rangle$'s, the probability that the
apparatus shows the value $o_k$ if ${\mathsf A}$ is registered on it in this
final state is $p^{\mathsf O}_\phi(o_k)$, which is what the probability
reproducibility requires. However, if the objectification requirement is to be
satisfied, two condition must be met:
\begin{description}
\item[(A)] The final state of the apparatus must the convex combination of the
form
\begin{equation}\label{gemengA} tr_{\mathcal S}[{\mathsf U}({\mathsf T}\otimes
{\mathsf T}_{\mathcal A}){\mathsf U}^\dagger] = \sum_j p^{\mathsf
O}_\phi(o_j)|\psi_j\rangle\langle\psi_j|\ .
\end{equation}
\item[(B)] The right-hand side of Eq.\ (\ref{gemengA}) must be the {\em
gemenge structure} of the state.
\end{description}

The notion of gemenge will play an important role in the reasoning of the
present paper. The term has been introduced in Ref.\ \cite{BLM}, some authors
(e.g., Ref.\ \cite{d'Espagnat}) use also the term 'proper mixture', Ref.\
\cite{ludwig1} calls it 'direct mixture'. The crucial point is that the convex
decomposition
\begin{equation}\label{defgem} {\mathsf T} = \sum_{k=1}^n w_k {\mathsf T}_k
\end{equation} of any state ${\mathsf T}$ (state operator) can be a gemenge
only if its preparation procedure ${\mathbf P}({\mathsf T})$ is a random
mixture with rates (frequencies) $w_k$ of preparations ${\mathbf P}({\mathsf
T}_k)$, where each ${\mathbf P}({\mathsf T}_k)$ is some preparation procedure
for ${\mathsf T}_k$, $k = 1,\cdots,n$. The preparation mixture can be done by
humans or result from some process in nature.

Thus, gemenge concerns a physical property of preparation rather than any
mathematical one of the right-hand side of Eq.\ (\ref{defgem}) (such as
${\mathsf T}_k$ being vector states or being mutually orthogonal, etc). From
the mathematical point of view, many different convex decompositions of a
general state ${\mathsf T}$ may exist. All possible components of such convex
combinations form a so-called 'face' in the space of state operators (cf.\
\cite{ludwig1}, P. 75). A state is 'extremal' if it lies in a zero-dimensional
face, that is, if it cannot be written as a non-trivial convex
combination. Extremal states are described by projections onto one-dimensional
subspaces of the Hilbert space. A preparation of ${\mathsf T}$ selects only
one of the mathematically possible convex decompositions of ${\mathsf T}$.

A random mixture of preparations is not uniquely determined by the preparation
process. It can be coarsened or refined, i.e., some of ${\mathbf P}({\mathsf
T}_k)$ can be combined into one preparation procedure or ${\mathbf P}({\mathsf
T}_k)$ for some $k$ can itself be a random mixture of other preparations.
\begin{df} The finest convex decomposition of state ${\mathsf T}$ defined by
its preparation as gemenge is called {\em gemenge structure} of ${\mathsf T}$.
\end{df} Thus, gemenge structure of ${\mathsf T}$ is uniquely determined by
its preparation. For extremal states, there is always only one gemenge
structure, the trivial one, independently of how it was prepared.

It may be advantageous to distinguish the mathematical convex combination of
states from their gemenge structure by writing the sum in Eq.\ (\ref{defgem})
as follows
\begin{equation}\label{defgem'} {\mathsf T} =
\left(\sum_{k=1}^n\right)_{\text{gs}} w_k {\mathsf T}_k
\end{equation} in the case that the right-hand side is a gemenge structure of
${\mathsf T}$.

The properties that follow directly from the definition of gemenge structure
and that will be needed later are described by the following theorem.
\begin{thm}
\begin{enumerate}
\item Gemenge structure is preserved by unitary dynamics,
$$
{\mathsf U}\left(\sum_k\right)_{\text{gs}} w_k {\mathsf T}_k{\mathsf
U}^\dagger = \left(\sum_k\right)_{\text{gs}} w_k {\mathsf U}{\mathsf
T}_k{\mathsf U}^\dagger\ :
$$
if the sum on the left-hand side describes a gemenge structure of ${\mathsf
T}$, then the gemenge structure of its evolution is described by the sum on
the right-hand side.
\item In the following sense, gemenge structure is also preserved by
composition of systems. Let ${\mathsf T}$ be a state of a composite system
${\mathcal S} + {\mathcal S}'$. The necessary and sufficient condition for the
partial trace over ${\mathcal S}'$ to have the gemenge structure described by
$$
tr_{{\mathcal S}'}[{\mathsf T}] = \left(\sum_k\right)_{\text{gs}} w_k {\mathsf
T}_k
$$
is that ${\mathsf T}$ itself has gemenge structure described by
$$
{\mathsf T} = \left(\sum_k\right)_{\text{gs}} w_k {\mathsf T}_k \otimes
{\mathsf T}'_k\ ,
$$
where ${\mathsf T}'_k$ are some states of ${\mathcal S}'$.
\end{enumerate}
\end{thm}

All these ideas on gemenges seem to be well known. Now, an important new point
will be added. In Ref.\ \cite{PHJT}, we have accepted the non-objectivity of
observables in its full extent, but we found a sufficient number of objective
properties of quantum systems elsewhere. The summary of the ideas can be
stated as follows
\begin{quote} {\bf Objectivity Assumption} A property is objective if its
value is uniquely determined by a preparation according to the rules of
standard quantum mechanics. The 'value' is the value of the mathematical
expression that describes the property and it may be more general than just a
real number. No registration is necessary to establish such a property but a
correct registration cannot disprove its value; in many cases, registrations
can confirm the value.
\end{quote} Objectivity Assumption led to a new realist interpretation of
quantum mechanics, see the extended discussion in Ref.\ \cite{PHJT}. It leads
also to a new meaning of gemenge structure: any individual system prepared in
the state (\ref{defgem'}) is objectively in one of the states ${\mathsf T}_k$,
because each of the systems has been prepared by one of the preparations
${\mathbf P}({\mathsf T}_k)$, and the probability that ${\mathbf P}({\mathsf
T}_k)$ has been used is $w_k$.

Let us return to our point (B), which can now be written as
$$
tr_{\mathcal S}[{\mathsf U}({\mathsf T}\otimes {\mathsf T}_{\mathcal
A}){\mathsf U}^\dagger] = \left(\sum_j\right)_{\text{gs}} p^{\mathsf
O}_\phi(o_j)|\psi_j\rangle\langle\psi_j|\ .
$$
According to the meaning of the gemenge structure, this equation expresses the
following property: after each registration, apparatus ${\mathcal A}$ is
objectively in one of the states $|\psi_j\rangle\langle\psi_j|$ and it is in
this state with probability $p^{\mathsf O}_\phi(o_j)$. This is exactly what
objectification requirement is meant to be. Thus, the two points (A) and (B)
can serve as an objectification criterion.

We can also understand why Beltrametti-Cassinelli-Lahti model of
premeasurement does not satisfy the objectification criterion. Indeed, the end
state ${\mathsf T}\otimes {\mathsf T}_{\mathcal A}$ of the system is ${\mathsf
U} (\phi \otimes \psi)$ (Eq.\ (\ref{finalSA})), which is a vector state and
can therefore have only a trivial gemenge structure. However, Point 2 of
Theorem 2 implies that this is not compatible with state $tr_{\mathcal
S}[{\mathsf U}({\mathsf T}\otimes {\mathsf T}_{\mathcal A}){\mathsf
U}^\dagger]$ being a non-trivial gemenge. Thus, we have shown a simple no-go
theorem. An analogous difficulty holds for more general models of
premeasurement described in Ref.\ \cite{BLM} and the book contains more
general no-go theorems. This is called {\em problem of objectification}. In
fact, our theorem and main idea of proof are similar to those given in Ref.\
\cite{BLM}. The rest of the paper will look for a reason why the vector state
${\mathsf U} (\phi \otimes \psi)$ must be replaced by a non-trivial gemenge so
that the objectification criterion can be satisfied.

\subsection{Repeatable premeasurement and von-Neumann model} In order to
define what a repeatable premeasurement is, we need the notion of state
transformer. To this aim, let us first calculate the final state of system
${\mathcal S}$ after a Beltrametti-Cassinelli-Lahti premeasurement is
finished:
$$
tr_{\mathcal A}[{\mathsf U}(|\phi\rangle\langle\phi| \otimes|
\psi\rangle\langle\psi|){\mathsf U}^\dagger] = \sum_k p^{\mathsf
O}_\phi(o_k)|\Phi_k\rangle\langle\Phi_k|\ .
$$
The part of the sum on the right-hand side corresponding to the result of
premeasurement lying in the set $X$ is
\begin{equation}\label{sttrans} {\mathcal I}(X)(|\phi\rangle\langle\phi|) =
\sum_{o_k\in X} p^{\mathsf O}_\phi(o_k)|\Phi_k\rangle\langle\Phi_k|\ .
\end{equation} The right-hand side is not a state, because it is not
normalised. Its trace is the probability that the result lies in $X$,
$$
p^{\mathsf O}_{\mathsf T}(X) = tr[{\mathcal I}(X)({\mathsf T})]
$$
if the initial state of ${\mathcal S}$ is ${\mathsf T}$. The quantity
${\mathcal I}(X)$ is an operation-valued measure and is called {\em state
transformer}. For more details, see Ref.\ \cite{BLM}.

\begin{df} A premeasurement is called {\em repeatable} if its state
transformer satisfies the equation
\begin{equation}\label{repeat} tr[{\mathcal I}(Y)({\mathcal I}(X)({\mathsf
T}))] = tr[{\mathcal I}(Y \cap X)({\mathsf T})]
\end{equation} for all subsets of possible values $X$ and $Y$ and all possible
states ${\mathsf T}$ of ${\mathcal S}$.
\end{df} That is, the repetition of the premeasurement on ${\mathcal S}$ does
not lead to any new result from the probabilistic point of view. To see
whether the state transformer (\ref{sttrans}) satisfies Eq. (\ref{repeat}),
let us rewrite it as follows:
$$
\sum_{o_k\in X} p^{\mathsf O}_\phi(o_k)|\Phi_k\rangle\langle\Phi_k| =
\sum_{o_k\in X}{\mathsf K}_k|\phi\rangle\langle\phi|{\mathsf K}^\dagger_k\ ,
$$
where
$$
{\mathsf K}_k = \sum_l|\varphi_{kl}\rangle\langle\phi_{kl}|\ .
$$
One can show that this relation is general,
$$
{\mathcal I}(X)({\mathsf T}) = \sum_{o_k\in X}{\mathsf K}_k{\mathsf T}{\mathsf
K}^\dagger_k\ ,
$$
for proof, see Ref.\ \cite{BLM}. We have then
$$
{\mathcal I}(Y)({\mathcal I}(X)({\mathsf T})) = \sum_{o_l\in X}{\mathsf
K}_l\left(\sum_{o_k\in X}{\mathsf K}_k{\mathsf T}{\mathsf
K}^\dagger_k\right){\mathsf K}^\dagger_l = \sum_{o_l\in X}\sum_{o_k\in
X}({\mathsf K}_l{\mathsf K}_k){\mathsf T}({\mathsf K}_l{\mathsf K}_k)^\dagger\
.
$$
Eq.\ (\ref{repeat}) would be satisfied if
\begin{equation}\label{K} {\mathsf K}_l{\mathsf K}_k = {\mathsf K}_k
\delta_{kl}\ ,
\end{equation} which is in general not the case.

Let us therefore restrict ourselves to measurement couplings satisfying
\begin{equation}\label{vNeum} \phi_{kl} = \varphi_{kl}\ .
\end{equation} This model is called von-Neumann premeasurement because it was
first described in Ref.\ \cite{JvN}\footnote{In fact, von-Neumann
premeasurement is slightly more general in the sense that it is a
premeasurement of a function $f({\mathsf O})$, where $f$ need not be
bijective, cf.\ Ref.\ \cite{BLM}.}.

For von-Neumann premeasurement, the operator ${\mathsf K}_k$ is the projection
${\mathsf E}^{\mathsf O}_k$ on the eigenspace of $o_k$,
$$
{\mathsf K}_k = \sum_l|\phi_{kl}\rangle\langle\phi_{kl}|
$$
and Eq.\ (\ref{K}) is satisfied. Thus, von-Neumann premeasurement is a special
case of repeatable premeasurement.

The vector states $\Phi_k$ given by Eq.\ (\ref{Phik}) are orthonormal for
von-Neumann premeasurements. Thus, the final state of the apparatus given by
Eq.\ (\ref{finalA}) reduces to (\ref{gemengA}) and Point (A) of our
objectification criterion is satisfied. As for Point (B), it is not satisfied
even for the more general Beltrametti-Cassinelli-Lahti model of
premeasurement. Hence, the objectification requirement does not hold for
von-Neumann premeasurements, and it is therefore not a measurement.

Von Neumann himself postulated that measurements define another, non-unitary
and indeterministic kind of evolution in which the state of ${\mathcal S}$
randomly jumps into one of the eigenstates of the measured observable (Ref.\
\cite{JvN}, PP. 217, 351). This was called {\em collapse of the wave function}
by Bohm (Ref.\ \cite{bohm}, P. 120).

\section{Comparison with real experiment. \\ Importance of detectors} The
theoretical models of the previous section ought to describe and explain at
least some aspects of real experiments. This section will try to go into all
experimental details that can be relevant to our theoretical understanding.

First, we briefly collect what we shall need about detectors. Microsystem
${\mathcal S}$ to be detected interacts with the sensitive matter of the
detector so that some part of energy of ${\mathcal S}$ is transferred to the
detector. Mostly, ${\mathcal S}$ interacts with many subsystems of the
sensitive matter exciting each of them because the excitation energy is much
smaller than the energy of ${\mathcal S}$. The resulting subsystem signals are
collected, or amplified and collected so that they can be distinguished from
noise. For example, in ionization detectors, many atoms or molecules of the
sensitive matter are turned into electron-ion pairs. If the energy of
${\mathcal S}$ is much higher than the energy of one ionisation, e.g.\ about
10 eV, then many electron-ion pairs are produced and the positive as well as
the negative total charge is collected at electrodes \cite{leo}.

In the so-called cryogenic detectors \cite{stefan}, ${\mathcal S}$ interacts,
e.g., with superheated superconducting granules by scattering off a nucleus
and the phase transition from the superconducting into the normally conducting
phase of only one granule leads to a perceptible electronic signal. A detector
can contain very many granules (typically $10^9$) in order to enhance the
probability of such scattering if the interaction between ${\mathcal S}$ and
the nuclei is very weak (WIMP, neutrino). Modern detectors are constructed so
that their signal is electronic. For example, to a scintillating film, a
photomultiplier is attached, etc., see Ref.\ \cite{leo}.

In any case, in order to make a detector respond ${\mathcal S}$ must loose
some of its energy to the detector. The larger the loss, the better the
signal. Thus, most detectors are built in such a way that ${\mathcal S}$
looses all its kinetic energy and is absorbed by the detector (in this way,
also its total momentum can be measured). Let us call such detectors {\em
absorbing}. If the bulk of the sensitive matter is not large enough,
${\mathcal S}$ can leave the detector after the interaction with it, in which
case we call the detector {\em non-absorbing}. Observe that a detector is
absorbing even if most copies of ${\mathcal S}$ leave the detector without
causing a response but cannot leave if there is a response (e.g., neutrino
detectors).

Suppose that ${\mathcal S}$ is prepared in such a way that it must cross a
detector. Then, the probability of the detector response is generally $\eta <
1$. We call a detector {\em ideal}, if $\eta = 1$.

An important assumption, corroborated by all experiments, is that a real
detector either gives a signal or remains silent in each individual
registration. This corresponds here to the objectification requirement.

After these preparatory remarks, consider a typical repeatable premeasurement
as described in textbooks (see, e.g., Ref.\ \cite{peres}, P. 27, where it is
called 'repeatable test'), for example a Stern-Gerlach-like measurement of
spin. A coordinate system $\{x^1,x^2,x^3\}$ is chosen. Silver atoms evaporate
in an oven $O$, form a beam $B_0$ along $x^2$-axis passing through a velocity
selector $S$, and then through an inhomogeneous magnetic field produced by
device $M_1$. $M_1$ splits $B_0$ into two beams, $B_{1+}$ and $B_{1-}$, of
which $B_{1+}$ is associated with positive and $B_{1-}$ with negative spin
$x^1$-component, the corresponding vector states being denoted by $|1+\rangle$
and $|1-\rangle$. Beam $B_{1-}$ is blocked off by a shield. This is the
preparatory part of the experiment.

Next, beam $B_{1+}$ runs through another magnetic device, $M_3^{(1)}$ with
centre at $\vec{x}_{(1)}$ and finally strike an array of ideal detectors
$\{D^{(1)}_k\}$ placed and oriented suitably with respect to $M_3^{(1)}$. Two
detectors of array $\{D^{(1)}_k\}$ respond, let us denote them by $D_+$ and
$D_-$, revealing the split of $B_{1+}$ into two beams, $B_{3+}$ and $B_{3-}$,
caused by $M_3^{(1)}$. Let the orientation of $M_3^{(1)}$ be such that
$B_{3+}$ corresponds to positive and $B_{3-}$ to negative spin
$x^3$-component, the states of silver atoms being $|3+\rangle$ or
$|3-\rangle$. The beams $B_{3+}$ and $B_{3-}$ are spatially sufficiently
separated so that their coordinates $\vec{x}_{3+}$ and $\vec{x}_{3-}$ at the
point where they strike the detectors can be considered as classical
values. In any case, they are measured by the detectors in a rather
coarse-grained way. Let us call experiment I what is performed by $O$, $S$,
$M_1$, $M_3^{(1)}$ and $\{D^{(1)}_k\}$.

Let us now remove $\{D^{(1)}_k\}$, place device $M_3^{(2)}$ of the same
macroscopic structure and orientation as $M_3^{(1)}$ with centre position
$\vec{x}_{(2)}$ in the way of $B_{3+}$ so that $B_{3-}$ passes by and arrange
array $\{D^{(2)}_k\}$ so that it has the same relative position with respect
to $M_3^{(2)}$ as $\{D^{(1)}_k\}$ had with respect to $M_3^{(1)}$. Now, only
one detector will respond, namely that at the position $\vec{x}_{3+} -
\vec{x}_{(1)} + \vec{x}_{(2)}$. Let us call experiment II what is performed by
$O$, $S$, $M_1$, $M_3^{(1)}$, $M_3^{(2)}$ and $\{D^{(2)}_k\}$. The result of
experiment II is described as 'two consecutive identical tests following each
other with a negligible time interval between them ... yield identical
outcomes' in Ref.\ \cite{peres}.

Clearly, experiment II does not consist of two copies of experiment I
performed after each other. The only repetition is that device $M_3^{(2)}$ is
placed after $M_3^{(1)}$ and has the same structure and orientation with
respect to its incoming beam $B_{3+}$ as $M_3^{(1)}$ has with respect to
$B_{1+}$. Device $M_3^{(1)}$ splits $B_{1+}$ into $B_{3+}$ and $B_{3-}$ but
$M_3^{(2)}$ does not split $B_{3+}$. One may say that it leaves $B_{3+}$
unchanged. Let us define the action of device $M_3^{(k)}$ together with the
{\em choice} of ($\pm$)-beam for each $k = 1,2$ as a test (in the sense of
Ref.\ \cite{peres}) or a premeasurements. Let the outcomes be the thought
response of an imaginary detector placed in the way of the chosen beam. Then
the (counterfactual) outcomes can be assumed to be identical indeed and we
have an example of repeatable premeasurement that satisfies Definition 1.

The procedures defined in this way are premeasurements that can be described
by von-Neumann model. The macroscopic positions $\vec{x}_{3+}$ or
$\vec{x}_{3-}$ of the atom after it passes the magnet can be considered as the
eigenvalues of the pointer observable associated with effects
$|3\pm\rangle\langle 3\pm|$. However, the premaesurement cannot be considered
as an instance of registration because it does not give us any information
about the silver atoms. Try to suppose, e.g., that the arrangement measures
effects $|3\pm\rangle\langle 3\pm|$ depending on which of the outgoing beams
is chosen. Now, how can we recognise whether the outcome is 'yes' or 'not'?
There is no change of a classical property of an apparatus due to its
interaction with a microsystem that would indicate which of the values
$\vec{x}_{3+}$ and $\vec{x}_{3-}$ results. But premeasurement is allowed not
to give definite responses by each individual action. To obtain definite
values, additional detectors are needed. Without the additional detector,
however, this real premeasurement is not a measurement.

Suppose next that there are non-absorbing ideal detectors that do not disturb
the spin state of the atom. This might work, at least approximately. Let
experiment I' be the same as I with the only change that the array
$\{D^{(1)}_k\}$ is replaced by $\{D^{p(1)}_k\}$ containing the non-absorbing
detectors. Let experiment II' starts as I' and proceeds as II but with
$\{D^{(2)}_k\}$ replaced by $\{D^{p(2)}_k\}$ made from the non-absorbing
detectors. Clearly, the action of $\bigl(M_3^{(j)} + \{D^{p(j)}_k\}\bigr)$ for
each $j=1,2$ is a repeatable premeasurement according to Definition 15, and it
is even a repeatable measurement because of the responses of the real
detectors, but it definitely cannot be described by a von-Neumann theoretical
model. For the detectors to response, some part of the energy of the atoms is
needed, so that condition (\ref{vNeum}) is not satisfied.

An interesting difference emerges here between what we can say about the
system (silver atom) on the one hand and about states on the other in their
relation to the beams $B_{3+}$ and $B_{3-}$. Whereas $B_{3+}$ is associated
with $|3+\rangle$ and $B_{3-}$ with $|3-\rangle$, each atom is in a linear
superposition of the two states $|3+\rangle$ and $|3-\rangle$ that equals to
the prepared state $|1+\rangle$. One can not even say that all atoms in beam
$B_{3+}$ are in state $|3+\rangle$ because no atom is just in $B_{3+}$. Unlike
the states, the atoms are not divided between the beams. Indeed, the two beams
could be guided so that no detectors are in their two ways and that they meet
each other again. Then, they would interfere and if the two ways are of equal
length, so that no relative phase shift results, the original state
$|1+\rangle$ would result. This would happen even if the beams are very thin,
containing always at most one silver atom. Hence, each atom had to go both
ways simultaneously.

Let us observe that each of the beams $B_{3+}$ and $B_{3-}$ by itself behave
as if it were a prepared beam of silver atoms in a known state, which is
$|3+\rangle$ and $|3-\rangle$, respectively. The voluntary element of beam
choice in this experiment can be interpreted neither as a preparation, nor as
a reselection of ensemble, nor as a collapse of the wave function. The fact
that we place some arrangement $A$ of devices that do not contain any detector
in the way of beam $B_{3+}$ and leave $B_{3-}$ alone justifies our use of
state in $|3+\rangle$ in all calculations of what will be the outcome after
arrangement $A$ is passed. However, the {\em whole} outcome will be a linear
superposition of states in each of the two beams at the time the upper beam
passes $A$. Only if we put any detector after $A$ or, for that matter, a
detector or just a shield into the way of $B_{3-}$, then something like a
collapse of the wave function can happen. The arrangement with the shield in
the way of $B_{3-}$ is a preparation of the vector state $|3+\rangle$.

The analysis of the present section motivates the following
generalisation. First, an arrangement of devices that acts in agreement with
von-Neumann model of premeasurement is neither a registration nor a
preparation apparatus. Second:
\begin{rl} Any registration apparatus for microsystems must contain at least
one detector and every reading of an apparatus value is a signal from a
detector.
\end{rl} If Rule 1 turns out not to be generally valid, then our theory of
quantum measurement will work at least for those many cases in which it is.

\section{Cluster separability} Quantum systems of the same type are
indistinguishable and this leads to entanglement. It seems then, that
experiments with one particle might be disturbed by another particle of the
same type, even if it were prepared independently, far away from the
first. One can avoid similar problems by adding some assumption of locality to
the axioms of quantum mechanics.

In the relativistic theory, one starts with the requirement that space-time
symmetries of an isolated system (i.e., that is alone in space) be realised by
unitary representations of Poincar\'{e} group on the Hilbert space of states,
see Refs.\ \cite{Weinberg} and \cite{haag}. Then, the {\em cluster
decomposition principle}, a locality assumption, states that if multi-particle
scattering experiments are studied in distant laboratories, then the
$S$-matrix element for the overall process factorizes into those concerning
only the experiments in the single laboratories. This ensures a factorisation
of the corresponding transition probabilities, so that an experiment in one
laboratory cannot influence the results obtained in another one. Cluster
decomposition principle implies non-trivial local properties of the theory
underlying the $S$-matrix, in particular it plays a crucial part in making
local field theory inevitable (cf. Ref.\ \cite{Weinberg}, Chap. 4).

In the phenomenological theory of relativistic or non-relativistic many-body
systems, Hilbert space of an isolated system must also carry a unitary
representation of Poincar\'{e} or Galilei group. Then, the so-called cluster
separability is a locality assumption, see, e.g., Refs.\ \cite{KP} or
\cite{coester} and references therein. It is a condition on interaction terms
in the generators of the space-time symmetry group saying: if the system is
separated into disjoint subsystems (=clusters) by a sufficiently large
spacelike separation, then each subsystem behaves as an isolated system with a
suitable representation of space-time symmetries on its Hilbert space, see
Ref.\ \cite{KP}, Sec.\ 6.1. Let's call this principle {\em cluster
separability I}.

Another special case of locality assumption has been described by Peres, Ref.\
\cite{peres}, p. 128. Let us reformulate it as follows
\par \vspace{.5cm}\noindent {\bf Cluster Separability II} No quantum
experiment with a system in a local laboratory is affected by the mere
presence of an identical system in remote parts of the universe.
\par \vspace{.5cm} \noindent It is well known (see, e.g., Ref.\ \cite{peres},
p. 136) that this principle leads to restrictions on possible statistics
(fermions, bosons). What is less well known is that it also motivates
non-trivial locality conditions on states that can be prepared and on
observables that can be registered.

The locality condition is formulated in Ref.\ \cite{peres}, p. 128:
\begin{quote} \mbox{...} a state $\mathsf w$ is called remote if $\|{\mathsf
A}{\mathsf w}\|$ is vanishingly small, for any operator ${\mathsf A}$ which
corresponds to a quantum test in a nearby location. ... We can now show that
the entanglement of a local quantum system with another system in a remote
state (as defined above) has no observable effect.
\end{quote} This is a condition on ${\mathsf A}$ inasmuch as there has to be
at least one remote state for ${\mathsf A}$.

However, Peres does not warn that the standard operators of quantum mechanics,
which are in fact generators of space-time symmetries, do not satisfy his
condition on ${\mathsf A}$. Similarly, basic observables of relativistic-field
or many-body theories are generators of Poincar\'{e} or Galilei groups and so
they do not satisfy the locality condition, either. It follows that cluster
separability II is logically independent from the cluster decomposition or of
cluster separability I. Of course, this does not mean that the basic
observables are to be rejected. They are very useful if the assumption of
isolated system is a good approximation. However, it is definitely a bad one
for quantum theory of measurement.

The present section expresses Peres' locality condition with the help of the
so-called $D$-local observables. Based on this analysis, it then introduces
the key notions of separation status and of its change. This is a modification
of standard quantum mechanics that leads to a possibility of prescribing new
rules for evolution of systems changing their separation status. Let us
explain everything, working in $Q$-representation of the common Hilbert space
${\mathbf H}$ and of operators on it, which will be represented by their
kernels. Then, one can also write tensor products as ordinary products and
indicate the order of factors by indices at system coordinates.

Suppose that vector state $\psi(\vec{x}_1)$ of particle 1 is prepared in our
laboratory as if no other particle of this type existed. Next, let vector
state $\phi(\vec{x}_2)$ of particle 2 of the same type be prepared
simultaneously in a remote laboratory. Then the state of the two particles
must be
\begin{equation}\label{symst} \Psi(\vec{x}_1,\vec{x}_2) =
\frac{1}{\sqrt{2}}\bigl(\psi(\vec{x}_1)\phi(\vec{x}_2) \pm
\phi(\vec{x}_1)\psi(\vec{x}_2)\bigr)
\end{equation} depending on the type statistics. If an observable with kernel
$a(\vec{x}_1;\vec{x}'_1)$ is now measured in our laboratory, it is equally
possible that the measurement is made on particle 1 or 2 and both can make a
contribution to the outcome. Hence, the correct observable is described by
two-particle kernel
\begin{equation}\label{symop} A(\vec{x}_1,\vec{x}_2;\vec{x}'_1,\vec{x}'_2) =
a(\vec{x}_1;\vec{x}'_1)\delta(\vec{x}_2-\vec{x}'_2) +
a(\vec{x}_2;\vec{x}'_2)\delta(\vec{x}_1-\vec{x}'_1)\ .
\end{equation}

In our language, the composite system of the two particles is an object but
each of the two particles is only a subobject. Thus, none of the particles
possesses its standard set of states and standard set of effects. There is
only a {\em common} one-particle Hilbert space, {\em common} standard set of
one-particle states and {\em common} standard set of one-particle effects that
the two particles share and that are formally equivalent to those of particle
1 if it were an object. These sets have only a formal, auxiliary
significance. From the common Hilbert space, the physical Hilbert space of the
composite system is formed by (anti)symmetrised tensor power containing states
such as (\ref{symst}). From the formal point of view,
$a(\vec{x}_1;\vec{x}'_1)$ (i.e., ${\mathsf a} \otimes {\mathsf 1}$) is not an
operator on the (anti)symmetrised Hilbert space, but the operator
(\ref{symop}) is. From the experimental point of view, the observable with
kernel $a(\vec{x}_1;\vec{x}'_1)$ is not measurable but that with kernel
(\ref{symop}) is.

There seems to be no control of states that are prepared anywhere in the world
and the different possibilities have different measurable consequences. For
example, the position of particle 1 as an object (i.e., without particle 2)
has kernel $a(\vec{x}_1;\vec{x}'_1) = \vec{x}_1\delta(\vec{x}_1-\vec{x}'_1)$
and suppose that the position is measured. Then, the average is
$$
\int d^3x_1\vec{x}_1\psi^*(\vec{x}_1)\psi(\vec{x}_1)\ .
$$
On the other hand, the existence of particle 2 leads to the average
\begin{multline*} \int d^3x_1d^3x_2
d^3x'_1d^3x'_2\Psi^*(\vec{x}'_1,\vec{x}'_2)
A(\vec{x}_1,\vec{x}_2;\vec{x}'_1,\vec{x}'_2) \Psi(\vec{x}_1,\vec{x}_2) \\ =
\int d^3x_1\vec{x}_1\psi^*(\vec{x}_1)\psi(\vec{x}_1) + \int d^3x_1\vec{x}_1
\phi^*(\vec{x}_1)\phi(\vec{x}_1)\ .
\end{multline*} The bigger the distance particle 2 has, the bigger the
difference is.

Cluster separability II can now be stated as follows. The change of ${\mathcal
S}_1$ state due to some actions in a remote laboratory would not be measurable
if the wave-function support of the remote system did not intersects domain
$D$ of the laboratory and if the observables that are measured were $D$-local
in the following sense.
\begin{df} Let $a(\vec{x}_1;\vec{x}'_1)$ be an observable of ${\mathcal S}_1$,
let $D$ be a domain of $\vec{x}_1$ and let
\begin{equation}\label{local1} \int d^3x_1a(\vec{x}_1;\vec{x}'_1)f(\vec{x}_1)
= \int d^3x'_1a(\vec{x}_1;\vec{x}'_1)f(\vec{x'}_1) = 0
\end{equation} if (supp\,$f)\cap D = \emptyset$, where $f$ is a test
function. Let us call such operators $D$-{\em local}.
\end{df}

Let us assume that (supp\,$\psi) \subset D$ and (supp\,$\phi) \cap D =
\emptyset$. If ${\mathcal S}_2$ has been prepared and the $D$-local kernel
$a_D(\vec{x}_1,\vec{x}'_1)$ is used instead of $a(\vec{x}_1;\vec{x}'_1)$ in
formula (\ref{symop}) defining operator ${\mathsf A}_D$ instead of ${\mathsf
A}$ and we obtain
\begin{multline*} \int_D
d^3x_1\int_Dd^3x'_1\int_Dd^3x_2\int_Dd^3x'_2\Psi^*(\vec{x}_1,\vec{x}_2)
A_D(\vec{x}_1,\vec{x}'_1;\vec{x}_2,\vec{x}'_2)\Psi(\vec{x}'_1,\vec{x}'_2) \\ =
\int_{-\infty}^\infty d^3x_1\int_{-\infty}^\infty
d^3x'_1\psi^*(\vec{x}_1)a(\vec{x}_1;\vec{x}'_1)\psi(\vec{x}'_1)
\end{multline*} as if no ${\mathcal S}_2$ existed. It follows that in this
case both rules for objects and rules for subobjects lead to the same results.

However, 'observables' that are usually associated with ${\mathcal S}_1$ are
not $D$-local. For example, the position operator violates the condition by
large margin, as seen above. In fact, the above analysis shows that such a
'position' is not measurable, be it represented by
$\vec{x}_1\delta(\vec{x}_1-\vec{x}'_1)$ or by
$\vec{x}_1\delta(\vec{x}_1-\vec{x}'_1) +
\vec{x}_2\delta(\vec{x}_2-\vec{x}'_2)$. Moreover, such an 'observable'
controls position of the system in the whole infinite space. This is utterly
different from observables that can be registered in a human
laboratory. Nevertheless, one can modify any observable by a map called
$\Lambda_D$ so that it becomes $D$-local and has the same averages in states
with supports in $D$ as the original observable had, see Appendix A.

It seems, however, that a similar problem exists even if particle 2 is not
remote: it can be prepared by a colleague on a neighbouring table in the same
laboratory. Still, the experience shows that measurements done on particle 1
on the first table are not disturbed by the activity on the second
table. Hence, the idea of cluster separability must work in the same way for a
less remote case, too.

But now the extent of the whole problem comes to light. For simple
microsystems, there are very many systems of the same type everywhere, at
least according to our realist interpretation of quantum mechanics. Clearly,
one could neglect the entanglement of a single microsystem ${\mathcal S}$ with
all microsystems of the same type, if ${\mathcal S}$ had a non-trivial {\em
separation status} in the following sense:
\begin{df} Let $D$ be a domain and system ${\mathcal S}$ be prepared in a
state with a $D$-local state operator ${\mathsf T}$. Let the probability to
register value of observable ${\mathsf E}(X)$ in set $X$ be $tr[{\mathsf
T}{\mathsf E}(X)]$ for any $D$-local observable ${\mathsf E}(X)$ of ${\mathcal
S}$. Then, domain $D$ is called separation status of ${\mathcal S}$.
\end{df} Here, ${\mathsf T}$ is a $D$-local state operator and ${\mathsf
E}(X)$ a $D$-local observable in the sense of Appendix A and the condition
means that the registration of ${\mathsf E}(X)$ is not disturbed by any state
different from ${\mathsf T}$. We can then view such microsystems as physical
objects.

For example, a microsystem that is alone in the Universe has separation status
$D = {\mathbb R}^3$. This is a form of the assumption of isolated
system. Measurable observables of such a system are the standard ones. The
same microsystem in a domain $D$ but which is surrounded by matter containing
a lot of microsystems of the same type such that supports of their states do
not intersect $D$ has separation status $D$ and its measurable observables are
the $D$-local ones. A trivial case of separation status for a microsystem is
if the only available modus of description for it is that of a subobject. This
has separation status $D = \emptyset$ and no observables of its own.

To formulate the idea of separation status mathematically, we allow an
exception to the rule for composition of identical systems. Let system
${\mathcal S}$ be prepared in the separation status $D$ and let ${\mathcal
S}'$ be a family of $N$ systems of the same type as ${\mathcal S}$ in a domain
$D'$, $D \cap D' = \emptyset$. Then the two systems ${\mathcal S}$ and
${\mathcal S}'$ are to be composed according to the rule for composition of
systems of different type. For example, let the wave function of ${\mathcal
S}$ be $\psi(\vec{x})$ and that of ${\mathcal S}'$ be
$\Psi(\vec{x}_1,\cdots,\vec{x}_N)$ that is symmetric or anti-symmetric in its
$N$ arguments according to the type. Then the wave function of composite
system ${\mathcal S} + {\mathcal S}'$ of $N+1$ subsystems of the same type
must be written as
\begin{equation}\label{firstst} \psi(\vec{x})\Psi(\vec{x}_1,\cdots,\vec{x}_N)\
.
\end{equation} Observe that wave function (\ref{firstst}) is not
(anti-)symmetric in all $N+1$ arguments! This is at variance with the formal
prescription dealing with families of identical systems. According to this
prescription, the wave function had to be
\begin{equation}\label{sectst}
\bigl[\psi(\vec{x})\Psi(\vec{x}_1,\cdots,\vec{x}_N)\bigr]_{s,a}\ ,
\end{equation} where the symbol $\bigl[\cdot \bigr]_{s,a}$ means
symmetrisation or anti-symmetrisations over all wave-function arguments
contained inside. This modification of standard quantum mechanics is essential
for our theory of measurement to work. Now, it also ought to be clear why we
do not employ Fock-space method to deal with identical systems: it
automatically (anti-)symmetrises over all systems of the same type.

The standard version of quantum mechanics as well as our interpretation
\cite{PHJT,hajicek} of it can be understood as a theory of systems with a
fixed status. Let us call these theories {\em fixed status quantum mechanics}
(FSQM). They deal with individual microsystems according to one set of rules
and with composite systems containing many particles of the same type
according to another set of rules. It neglects the obvious relations between
the two that make such an approach in principle inconsistent. However, the
method seems to work and the justification why it approximately works is the
cluster separability. It follows that FSQM has limits and that the limits have
to do with the cases when separation status of system $\mathcal S$
changes. The main idea of the present paper is that there is certain freedom
in choosing the state of $\mathcal S$ that results from a change of status
(see Sec.\ 5).

The simplest example of separation status change is as follows. Suppose that
wave function (\ref{firstst}) is evolved further by the some first-quantised
Hamiltonian according to prescriptions of standard quantum mechanics so that
the support of wave function $\psi(\vec{x})$ changes from $D$ to $D'$ (i.e.,
probability to find system $\mathcal S$ outside $D'$ is then negligible) while
$\Psi(\vec{x}_1,\cdots,\vec{x}_N)$ remains in $D'$. Thus, the separation
status of ${\mathcal S}$ becomes $\emptyset$ and ${\mathcal S}$ itself becomes
a subobject. One possibility for the resulting state will now be described by
(\ref{sectst}), where the wave functions are replaced by their evolved
versions. Observe that the change from state (\ref{firstst}) to (\ref{sectst})
is not unitary. This is in agreement with the fact that the set of observables
measurable on $\mathcal S$ was radically reduced.

Let us close this section by a brief remark on macroscopic systems. In
general, a macroscopic system ${\mathcal A}$ is a composite quantum system
with very many different microsystem constituents. One can subdivide these
microsystems into type classes. If we apply the basic rules of observable
construction for systems of identical microsystems, then e.g.\ the position
and momentum of any individual microsystem are not observables of ${\mathcal
A}$. However, depending on how large the considered microsystem is and on the
supports of all relevant states, some constituent microsystems can be
considered as approximately separated. In general, to construct measurable
observables for ${\mathcal A}$ is a non-trivial problem. For instance,
eigenvalues of energy are not measurable (the spectrum of any macroscopic
system is too dense for that). Instead, the average value of energy with some
variance is measurable, etc., see Ref.\ \cite{hajicek}. Or, $X$-rays can be
scattered by a crystal and so relative positions of its nuclei can be
recognised. But rather than a position of an individual nucleus it is a space
dependence of the average nuclear density due to all nuclei that is measured
by the scattering.

\section{Gemenge structure of final detector states} Sec.\ 3 motivated the
idea that the reading of a registration apparatus for microsystems is in fact
an electronic signal from a detector. This gives us much clearer notion of
registration apparatus. The main idea of Sec.\ 4 is that FSQM description of
microsystems has its limits. This consequence of basic assumptions of standard
quantum mechanics about indistinguishable microsystems leads to a significant
modification of quantum theory of measurement. The necessary changes are:
\begin{enumerate}
\item Each preparation of microsystem ${\mathcal S}$ must separate the
microsystem. Prepared state ${\mathsf T}$ must be $D$-local in a suitable
domain $D$.
\item Microsystem ${\mathcal S}$ can then be manipulated and controlled by
devices within $D$ such as electric and magnetic fields, matter shields,
detectors, etc.
\item Let macrosystem ${\mathcal A}$ such as a blocking shield, a scattering
target or a detector that contains microsystems indistinguishable from
${\mathcal S}$ lie in $D$. Corrections to FSQM description of the behaviour of
the composed system ${\mathcal S} + {\mathcal A}$ due to a possible separation
status change of ${\mathcal S}$ must be carefully chosen.
\end{enumerate} The usual method of FSQM is to specify initial states of both
${\mathcal S}$ and ${\mathcal A}$ before their interaction, choose some
appropriate interaction Hamiltonian and calculate the corresponding unitary
evolution of the composed system ${\mathcal S} + {\mathcal A}$ ignoring the
problem with separation status change. As shown in Sec.\ 2, the results are
wrong for registration apparatuses. We shall now try to choose some
corrections.

Let ${\mathcal S}$ be the registered microsystem and ${\mathcal A}$ be an
array of $N$ ideal monoatomic-gas ionisation detectors similar to that of
Sec. 3. Let index $k$ enumerate the detectors and let each detector be treated
as a system of identical atoms. Let each atom be modelled by a particle with
mass $\mu$, spin zero and a further degree of freedom, ionisation, with two
values, non-ionised and ionised. We simplify the model further by assuming
that the ionisation and translation degrees of freedom can be separated from
each other in such a way that they define two different formal subsystems,
${\mathcal A}_{\text{ion}}$ and ${\mathcal A}_{\text{tra}}$ of the whole real
macroscopic system ${\mathcal A}$. Let $\chi_{kn}$ be the state describing $n$
ionised atoms in $k$th detector. The states
$$
\prod_k \otimes \chi_{kn(k)}
$$
for all $n(k)$'s form a basis of the Hilbert space of ${\mathcal
A}_{\text{ion}}$, where $n(k)$ is a map of $\{1,\cdots,N\}$ into non-negative
integers. Let us assume that the initial state of ${\mathcal A}_{\text{ion}}$
is
$$
\psi = \prod_k \otimes \chi_{k0}\ ,
$$
the perfectly non-ionised state. We can further assume that the initial state
${\mathsf T}_{\text{tra}}$ of ${\mathcal A}_{\text{tra}}$ is close to maximum
entropy one with sufficiently low temperature so that ionisations due to
atomic collisions have a very low probability.

The measurement coupling is a coupling between ${\mathcal S}$ and the
ionisation degree of freedom of each atom in the sensitive matter of the whole
array. That is, ${\mathcal S}$ interacts directly only with ${\mathcal
A}_{\text{ion}}$. In a single detector, after the ionisation of the first
atom, all subsequent ionisations lie along a ray track inside the same
detector. An explanation of the fact that e.g.\ a spherical wave can produce a
straight track is given in Ref.\ \cite{mott}, where it is shown that the
position of the track head, the first ionisation of the track, determines the
track. This can be considered as a necessary property of every measurement
coupling that is possible in the case considered here. Let the measurement
coupling be that of the Beltrametti-Cassinelli-Lahti model, satisfying Eq.\
(\ref{unitar}), where
$$
\psi_k = \left(\prod_{j=1}^{k-1} \otimes \chi_{j0}\right)\otimes \left(\sum_n
a_n\chi_{kn}\right)\otimes \left(\prod_{j=k+1}^N \otimes \chi_{j0}\right)
$$
and $a_n$ are coefficients independent of $k$ satisfying $\sum_n |a_n|^2 =
1$. This is again a simplifying assumption: each ${\mathcal S}$ creates always
the same ionisation state in each detector.

In Sec.\ 2, states $\psi_k$ were called 'end states' of ${\mathcal A}$ and
they were eigenstates of observable ${\mathsf A}$ called 'pointer observable'.
Here, we prefer $\psi_k$ to be called {\em trigger states} because there is a
further evolution of ${\mathcal A}$ independent of ${\mathcal S}$ that leads
from $\psi_k$ to the concentration of charges at the electrodes, and an
electronic signal, of $k$th detector. This is due to a coupling between
${\mathsf A}_{\text{ion}}$ and ${\mathsf A}_{\text{tra}}$ mediated by the
electrostatic field of the electrodes: ionised atoms move in a different way
than the non-ionised ones. This motion leads to atom collisions and further
ionisation in a complicated irreversible process. Only then, the true end
states with true pointer values are achieved. There is no pointer observable,
the pointer values being some averages with some variances, in agreement with
the expectation of Refs.\ \cite{PHJT,hajicek}. However, what is important for
us happens already at the trigger stage and we can ignore the evolution from a
trigger state to a detector signal.

From the requirement that the measurement yields a definite result, an
important statement follows (cf.\ Sec.\ 2):
\begin{thm} A measurement coupling of a true registration must be such that
the end states $\varphi_{kl}$ of ${\mathcal S}$ are orthonormal,
\begin{equation}\label{endorth} \langle \varphi_{kl}|\varphi_{mn}\rangle =
\delta_{km}\delta_{ln}\ .
\end{equation}
\end{thm} The unitary evolution defined by the measurement coupling yields a
trigger state of the whole system ${\mathcal S} + {\mathcal A}_{\text{ion}}$
given by Eqs.\ (\ref{finalSA}). Then, the trigger state of ${\mathcal
A}_{\text{ion}}$, obtained from Eq.\ (\ref{finalA}) and (\ref{endorth}), is
given by Eq.\ (\ref{gemengA}).

According to Theorem 2, state (\ref{gemengA}) of ${\mathcal A}_{\text{ion}}$
has not the gemenge structure given by the right-hand side of Eq.\
(\ref{gemengA}) because of the entanglement with ${\mathcal S}$ due to state
(\ref{finalSA}). The reason is that state (\ref{finalSA}) contains much more
correlations between observables of ${\mathcal S}$ and ${\mathcal
A}_{\text{ion}}$ than just correlations between the states $\Phi_k$ and
$\psi_k$. To measure any of these correlations, we would always need some
observables of ${\mathcal S}$ that do not commute with ${\mathsf O}$ (see
Appendix B).

However, the assumption that the trigger state ${\mathcal A}_{\text{ion}}$ is
(\ref{finalSA}) seems to be an illusion. Microsystem ${\mathcal S}$ is
somewhere inside ${\mathcal A}$ at this stage and has become indistinguishable
from other microsystems of the same type within ${\mathcal A}$. There is
always a lot of them, either because they are present in the detectors before
the registration started or because the detector becomes quickly polluted by
them afterwards. Thus, the separation status of the system ${\mathcal S}$ has
changed from an object to a subobject and with it also the separation status
of the whole composite system ${\mathcal S} + {\mathcal A}$ has. The
applications of FSQM to two systems of different separation status is
different. In our case, system ${\mathcal S} + {\mathcal A}$ before the
interaction is a composite one and each of the subsystems is an object having
its states and observables. During and after the interaction, however,
${\mathcal S}$ ceases to be an object, becomes a part of ${\mathcal A}$ and
looses all of its observables except of ${\mathsf O}$. This is a deeper change
than just a change of state. Hence, the existence of most correlations that
are the content of state (\ref{finalSA}) is lost. The point is not that some
observables are difficult to measure but rather that these observables do not
exist at all. The only correlations that can remain are those between the
trigger states $\psi_k$ of ${\mathcal A}_{\text{ion}}$ and $\Phi_k$ of the
microsystem. They are the content of the state
$$
\sum_k |c_k|^2|\Phi_k\rangle \langle\Phi_k| \otimes |\psi_k\rangle
\langle\psi_k|\ .
$$
This motivates the following assumption:
\begin{rl} Let a microsystem ${\mathcal S}$ be detected by a detector
${\mathcal A}$ and the measurement coupling satisfy Eq.\ (\ref{endorth}) so
that the corresponding unitary evolution leads to the state (\ref{finalSA})
with ${\mathcal S}$ inside ${\mathcal A}$. Then, instead of (\ref{finalSA}),
the true state of ${\mathcal S} + {\mathcal A}_{\text{ion}}$ is
\begin{equation}\label{truend} \left(\sum_k\right)_{\text{gs}}
|c_k|^2|\Phi_k\rangle \langle\Phi_k| \otimes |\psi_k\rangle \langle\psi_k|\ .
\end{equation}

\end{rl} It then follows from Theorem 2 that the trigger state of ${\mathcal
A}_{\text{ion}}$ is
\begin{equation}\label{gemengA'} tr_{\mathcal S}[{\mathsf U}({\mathsf
T}\otimes {\mathsf T}_{\mathcal A}){\mathsf U}^\dagger] =
\left(\sum_j\right)_{\text{gs}} p^{\mathsf
O}_\phi(o_j)|\psi_j\rangle\langle\psi_j|\ .
\end{equation}

The content of Rule 2 is that only the correlations between the states
$\psi_k$ of ${\mathcal A}_{\text{ion}}$ and $\Phi_k$ of the microsystem
survive and all other correlations between ${\mathcal A}_{\text{ion}}$ and
${\mathcal S}$ are erased during the change of separation status of ${\mathcal
S} + {\mathcal A}$. What survives and what is erased is uniquely determined by
the Beltrametti-Cassinelli-Lahti model. In particular, the probability
reproducibility condition determines states $\varphi_{kl}$ from the initial
state $\psi$ of ${\mathcal A}_{\text{ion}}$ uniquely and the initial state
$\phi$ of ${\mathcal S}$ determines states $\Phi_k$ uniquely. Thus, the
additional evolution from state (\ref{finalSA}) to state (\ref{truend}) is
non-unitary but still deterministic. Rule 2 is a new basic assumption which
has to be added to quantum mechanics. To choose such an assumption, we have to
look at observations and experiments. Rule 2 is in an agreement with what is
observed.

A correct interpretation of Rule 2 distinguishes two cases. If the detectors
are absorbing, then states $\Phi_k$ in Eq.\ (\ref{truend}) ought to be
(anti-)symmetrised with states of other systems indistinguishable from
$\mathcal S$ within the $k$-th detector as in Eq.\ (\ref{sectst}). The
expression $|\Phi_k\rangle \langle\Phi_k|$ in it just symbolises the fact that
system ${\mathcal S}$ has been lost in the $k$-th detector. If they are
non-absorbing, then state (\ref{truend}) contains states $\psi_k $ leading to
detector signals on the one hand and describes the release of ${\mathcal S}$
in state $\Phi_k$ that is correlated with detector signals on the other.  Each
release is understood as an instance of preparation and the whole procedure is
a random mixture of these single preparations. In both cases, the end state of
${\mathcal A}_{\text{ion}}$ is (\ref{gemengA'}).

One can wonder whether a more detailed quantum mechanical model of what
happens during a change of separation status can be constructed. The reason
why this cannot be done within FSQM is that FSQM is not applicable to changes
of separation status. Hence, a new law added to FSQM is needed.

As an example of a system of non-absorbing detectors, the MWPC telescope for
particle tracking can be mentioned \cite{leo}. It is a stack of the so-called
multiwire proportional chambers (MWPC), which is arranged so that a particle
runs through exciting each of them. The resulting system of electronic signals
contains the information about the particle track.

A registration by a non-absorbing detector is similar to a scattering of a
microsystem by a macroscopic target. First, let us consider no-entanglement
processes such as the scattering of electrons on a crystal of graphite with an
interference pattern as a result \cite{DG} or the splitting of a laser beam by
a down-conversion process in a crystal of KNbO$_3$ (see, e.g., Ref.\
\cite{MW}). No-entanglement processes can be described by the following model.
Let the initial state of the target ${\mathcal A}$ be $\mathsf T$ and that of
the microsystem be $\phi$. We assume that the end state of the target is
${\mathsf T}'$ and the end-state of the microsystem is $\varphi$ and that we
have a unitary evolution:
$$
|\phi\rangle\langle\phi| \otimes {\mathsf T} \mapsto
|\varphi\rangle\langle\varphi| \otimes {\mathsf T}'\ .
$$
There is no entanglement of the two systems due to the interaction and there
is no necessity to divide the resulting correlations between ${\mathcal S}$
and ${\mathcal A}$ in what survives and what is erased. The end state is
already of the form (\ref{truend}) and it has a trivial gemenge structure. In
this way, our corrections of FSQM become trivial in this case.

A more complicated case is an entanglement scattering. Let microsystem
${\mathcal S}$ in initial state $\phi$ be scattered by a macrosystem
${\mathcal A}$ in initial state ${\mathsf T}$ and let this lead to excitation
of different microscopic subsystems ${\mathcal S}'_k$ of ${\mathcal A}$.
Scattering of neutrons on spin waves in ferromagnets, transmutation of nuclei
inside ${\mathcal A}$ or, for that matter, ionising an atom in a gas detector
are examples. We have, therefore, a more general situation than that in which
Rule 2 gives a unique result. It seems that the change of status must lead to
some correlations between ${\mathcal S}$ and ${\mathcal A}$ surviving and some
being erased. However, in this situation it must yet be investigated which is
which. Clearly, the definitive general rule must depend on the two interacting
systems and on the interaction Hamiltonian. More theoretical and experimental
work is necessary to guess the general rule.

\section{Conclusion} The present paper proposes some ideas based on cluster
separability with the aim to solve the objectification problem of quantum
measurement. Its main purpose is to show how the ideas work by studying
well-understood, restricted class of physical conditions in which the
following assumptions are a good approximation: (a) non-relativistic quantum
mechanics, (b) measurement performed directly on microsystems, (c)
Beltrametti-Cassinelli-Lahti model of measurement and (d) pointer readings
being signals from detectors.

Ref.\ \cite{BLM} defines and analyses the problem of objectification and shows
its insolubility: no-go theorems such as Theorem 6.2.1, P. 76. One of the
premises of all theorems of this kind is that standard quantum mechanics
(without any further assumptions such as that of collapse of the wave
function) is applicable to preparation and registration processes. The present
paper gives a physical justification of why this premise is not valid: during
preparation and registration, the system changes its separation status and
standard quantum mechanics 1) is not applicable to, and 2) does not contain
any rules for, such kind of evolution. Thus, new rules that govern changes of
separation status can be added without any contradiction with standard quantum
mechanics or proofs of no-go theorems. Rule 2 is an example of such a new
rule. Thus, the no-go theorem of Sec.\ 2 is avoided.

An important result of the present paper together with Refs.\
\cite{PHJT,hajicek} is a strongly improved understanding of preparation
procedure. First, any preparation gives the prepared system its objective
quantum properties such as states, gemenge structures, averages and variances
of observables etc. so that it is justified to speak of a physical
object. This is what we have called quantum object. Second, in certain sense,
a preparation must separate a microsystem from the set of identical
microsystems, at least approximately. Only then, it can be viewed as an
individual system and the standard notion of observable becomes applicable to
it. This is justified by the idea of cluster-separability. Third, a
preparation must isolate the microsystem so that it can be individually
manipulated by e.g.\ external fields or mater shields and registered by
detectors.

One trend in the post-Everett theory of quantum measurement is to avoid the
assumption of collapse of the wave function during registrations. In a sense,
the present paper is heading in the opposite direction. We even replace the
collapse by a more radical transformation, a change in microsystem description
including state spaces and observable algebras. This change is, in plain
words, a kind of loss of a registered object during its registration. However,
our result for non-absorbing detectors and the old idea by von Neumann have
some features in common.

After having shown that our ideas work under the simplified conditions listed
above we can start thinking about extending the method to more general
conditions. There is a lot of work to be done yet. First, we must turn to
other models of measurement, for example to different (non-ideal) kinds of
detectors or to the more realistic premeasurement models within the
non-relativistic quantum mechanics. The main point is again that the state
resulting from the evolution contains information about properties of the
composite system ${\mathcal S} + {\mathcal A}$ that could be measured only if
more observables than the registered one of ${\mathcal S}$ existed. Thus, a
change of this illusory state analogous to that given by Rule 2 could be
justified. In such a way, all no-go theorems could be defused. The exact
division line between correlations that survive and those that are erased
during the registrations and other processes might again be determined by a
careful analysis of observational facts.

Next, relativistic corrections have been neglected so that all notions and
rules of non-relativistic quantum mechanics could be used. An extension of the
present results to relativistic fields seems to be a realistic project because
cluster separability is valid in this field.

\section*{Appendix A: Construction of $D$-local observables} For the
construction, we need more mathematics. Let ${\mathcal L}_r({\mathcal H})$
denote the set of all self-adjoint operators on the Hilbert space ${\mathcal
H}$ that are bounded in the norm
\begin{equation}\label{opnorm} \|{\mathsf A}\| = \sup_{\|\psi\| = 1}\|{\mathsf
A}\psi\|\ .
\end{equation}

An operator ${\mathsf A}\in{\mathcal L}_r({\mathcal H})$ is positive,
${\mathsf A}\geq {\mathsf 0}$, where ${\mathsf 0}$ is the null operator, if
$$
\langle\phi|{\mathsf A}\phi\rangle \geq 0
$$
for all vectors $\phi\in{\mathcal H}$. The relation ${\mathsf A}\geq {\mathsf
B}$ defined by
$$
{\mathsf A}-{\mathsf B}\geq {\mathsf 0}
$$
is an ordering on this space. With this (partial) order relation, ${\mathcal
L}_r({\mathcal H})$ is an ordered Banach space.

\begin{df} Let ${\mathcal F}$ be the Boolean lattice of all Borel subsets of
${\mathbb R}^n$. A {\em positive operator valued (POV) measure}
$$
{\mathsf E} : {\mathcal F} \mapsto {\mathcal L}_r({\mathcal H})
$$
is defined by the properties
\begin{enumerate}
\item positivity: ${\mathsf E}(X) \geq 0$ for all $X\in {\mathcal F}\ ,$
\item $\sigma$-additivity: if $\{X_k\}$ is a countable collection of disjoint
sets in ${\mathcal F}$ then
$$
{\mathsf E}(\cup_k X_k) = \sum_k {\mathsf E}(X_k)\ ,
$$
where the series converges in weak operator topology, i.e., averages in any
state converge to an average in the state.
\item normalisation:
$$
{\mathsf E}({\mathbb R}^n) = {\mathsf 1}\ ,
$$
where ${\mathsf 1}$ is the identity operator on ${\mathcal H}$.
\end{enumerate} The number $n$ is called {\em dimension} of ${\mathsf E}$. The
operators ${\mathsf E}(X)$ for $X\in {\mathcal F}$ are called {\em effects}.
\end{df}

We denote by ${\mathcal L}_r({\mathcal H})^+_{\leq 1}$ the set of all effects.
\begin{thm} ${\mathcal L}_r({\mathcal H})^+_{\leq 1}$ is the set of elements
of ${\mathcal L}_r({\mathcal H})$ satisfying the inequality
\begin{equation}\label{effect} {\mathsf 0}\leq {\mathsf E}(X)\leq {\mathsf I}\
.
\end{equation}
\end{thm} For the proof, see Ref.\ \cite{ludwig1}.

A special case of POV measure is projection valued measure (PV measure). All
effects of a PV measure are projections onto subspaces of ${\mathcal H}$. The
spectral measure of a s.a. operator is a PV measure, hence POV measure is a
generalisation of a s.a. operator. More about POV measures as well as the
motivation for viewing them a quantum-mechanical observables, see Refs.\
\cite{ludwig1,BLM,peres}.

Let us denote by ${\mathcal H}_D$ the Hilbert space obtained by completion of
$C^\infty$-functions with support in $D$ with respect to the scalar product of
${\mathcal H}$. ${\mathcal H}_D$ is a closed linear subspace of ${\mathcal
H}$. Let ${\mathsf P}_D$ be the projection from ${\mathcal H}$ onto ${\mathcal
H}_D$.
\begin{df} Let
$$
\Lambda_D : {\mathcal L}_r({\mathcal H}) \mapsto {\mathcal L}_r({\mathcal H})
$$
be defined by
$$
\Lambda_D({\mathsf A}) = {\mathsf P}_D{\mathsf A}{\mathsf P}_D\ .
$$
Mapping $\Lambda_D$ is called $D$-{\em localization}.
\end{df} Clearly, $D$-localisation of any operator in ${\mathcal
L}_r({\mathcal H})$ is $D$-local. Everything that is measurable within $D$ can
be described by $D$-local observables. Of course, the $D$-localisation is not
a unitary map. For example, it does not preserve operator norm,
$$
\|\Lambda_D({\mathsf A})\| \leq \|{\mathsf A}\| .
$$

The operators and their $D$-localisations are considered as acting on
${\mathcal H}$. $D$-local operators leave ${\mathcal H}_D$ invariant and
define, therefore, also operators on Hilbert space ${\mathcal H}_D$.

We can use these facts in a construction of $D$-local POV measure on
${\mathcal H}_D$ from any observable ${\mathsf E}$ on ${\mathcal H}$ by
$D$-localising the effects ${\mathsf E}(X)$. The normalisation condition
becomes:
$$
\Lambda_D({\mathsf E}({\mathbb R}^n)) = {\mathsf P}_D{\mathsf 1}{\mathsf P}_D
= {\mathsf 1}_D\ ,
$$
where ${\mathsf 1}_D$ is the identity operator on ${\mathcal H}_D$.  Of
course, $D$-localisation of a projection will not be a projection in general
and so a $D$-localisation of a PV measure need not be a PV measure. Let us
call this construction $D$-localisation of POV measures. All $D$-local POV
measures commute with spectral projections of PV measure ${\mathsf
E}^{\vec{\mathsf Q}}(X)$, if $X \cap D =\emptyset$. ${\mathsf E}^{\vec{\mathsf
Q}}(X)$ is the spectral measure of the position operator $\vec{\mathsf
Q}$. Thus, the restriction to $D$-local observables may be formally understood
as super\-selection rules.

Everything can be easily extended from vector to general states; the state
operators must just be $D$-local. If the map $\Lambda_D$ is involved in their
construction it must be followed by a suitable normalisation.

\section*{Appendix B: Complete set of correlations \\ in a vector state of a
composite system} Consider a composite system with constituents ${\mathcal S}$
and ${\mathcal S}'$ in vector state
\begin{equation}\label{schmidt} \Phi = \sum_k c_k \phi_k \otimes \phi'_k\ ,
\end{equation} $\{\phi_k\}$ being a basis of ${\mathcal H}_{\mathcal S}$,
$\{\phi'_k\}$ that of ${\mathcal H}_{{\mathcal S}'}$ and $c_k$ satisfying
$$
\sum_k |c_k|^2 = 1\ .
$$
In fact, any vector state of ${\mathcal S} + {\mathcal S}'$ can be written in
the form (\ref{schmidt}), which is called Schmidt decomposition (see, e.g.,
\cite{peres}, P. 123).

Let ${\mathsf O}$ be an observable of ${\mathcal S}$ and ${\mathsf O}'$ of
${\mathcal S}'$ and let us introduce the following abbreviations:
\begin{eqnarray*} \langle {\mathsf O} \rangle_\Phi &=& \langle \Phi|{\mathsf
O} \otimes {\mathsf 1} |\Phi\rangle\ ,\\ \langle {\mathsf O}' \rangle_\Phi &=&
\langle \Phi|{\mathsf 1} \otimes {\mathsf O}' |\Phi\rangle\ ,\\ \langle
{\mathsf O}{\mathsf O}' \rangle_\Phi &=& \langle \Phi|{\mathsf O} \otimes
{\mathsf O}' |\Phi\rangle\ ,\\ \Delta_\Phi {\mathsf O} &=& \sqrt{\langle
{\mathsf O}^2 \rangle_\Phi - \langle {\mathsf O} \rangle_\Phi^2}\ ,\\
\Delta_\Phi {\mathsf O}' &=& \sqrt{\langle {\mathsf O}^{\prime 2} \rangle_\Phi
- \langle {\mathsf O}' \rangle_\Phi^2}\ .
\end{eqnarray*}

The normalised correlation of ${\mathsf O}$ and ${\mathsf O}'$ in $\Phi$ is
defined by
\begin{equation}\label{correl} \rho({\mathsf O},{\mathsf O}',\Phi) =
\frac{\langle {\mathsf O}{\mathsf O}' \rangle_\Phi - \langle {\mathsf O}
\rangle_\Phi\langle {\mathsf O}' \rangle_\Phi}{\Delta_\Phi {\mathsf
O}\Delta_\Phi {\mathsf O}'}\ .
\end{equation} The normalised correlation always satisfies
$$
-1 \leqq \rho({\mathsf O},{\mathsf O}',\Phi) \leqq 1
$$
because of Schwarz' inequality. If $\rho({\mathsf O},{\mathsf O}',\Phi) = 0$
observables ${\mathsf O}$ and ${\mathsf O}'$ are uncorrelated, if
$\rho({\mathsf O},{\mathsf O}',\Phi) = \pm 1$ they are strongly
correlated/anti-correlated.

Let us first apply these formulae to projections,
$$
{\mathsf P}_k = |\phi_k\rangle \langle \phi_k|\ ,\quad {\mathsf P}'_k =
|\phi'_k\rangle \langle \phi'_k|\ .
$$
Simple calculations yield
$$
\langle {\mathsf P}_k \rangle_\Phi = \langle {\mathsf P}'_k \rangle_\Phi =
|c_k|^2\ ,
$$
$$
\Delta_\Phi {\mathsf P}_k = \Delta_\Phi {\mathsf P}'_k = |c_k|\sqrt{1 -
|c_k|^2}\ ,
$$
$$
\langle {\mathsf P}_k {\mathsf P}'_l\rangle_\Phi = |c_k|^2\delta_{kl}\ .
$$
Thus,
$$
\rho({\mathsf P}_k, {\mathsf P}'_k,\Phi) = 1\ .
$$
It follows that ${\mathsf P}_k$ and ${\mathsf P}'_k$ are strongly correlated
in $\Phi$.

Next, consider bounded, s.a.\ operators
\begin{eqnarray*} {\mathsf P}_{\alpha kl} &=& e^{i\alpha}|\phi_k\rangle
\langle \phi_l| + e^{-i\alpha}|\phi_l\rangle \langle \phi_k|\ ,\\ {\mathsf
P}'_{\alpha kl} &=& e^{i\alpha}|\phi'_k\rangle \langle \phi'_l| +
e^{-i\alpha}|\phi'_l\rangle \langle \phi'_k|
\end{eqnarray*} for $k \neq l$. We calculate:
$$
\langle {\mathsf P}_{\alpha kl} \rangle_\Phi = \langle {\mathsf P}'_{\alpha
kl} \rangle_\Phi = 0\ ,
$$
$$
\Delta_\Phi {\mathsf P}_{\alpha kl} = \Delta_\Phi {\mathsf P}'_{\alpha kl} =
\sqrt{|c_k|^2 + |c_l|^2}\ ,
$$
$$
\langle {\mathsf P}_{\alpha kl} {\mathsf P}'_{\beta kl}\rangle_\Phi =
e^{i(\alpha+\beta)}c^*_k c_l + e^{-i(\alpha+\beta)}c^*_l c_k\ .
$$
Thus,
$$
\rho({\mathsf P}_{\alpha kl}, {\mathsf P}'_{\beta kl},\Phi) =
\frac{e^{i(\alpha+\beta)}c^*_k c_l + e^{-i(\alpha+\beta)}c^*_l c_k}{|c_k|^2 +
|c_l|^2}\ .
$$
It follows that correlations of the observables ${\mathsf P}_{\alpha kl}$ and
$ {\mathsf P}'_{\beta kl}$ in state $\Phi$ contain complete information about
all coefficient $c_k$ except for their common phase and so determine state
$\Phi$. It is sufficient to use just two choices of $\alpha$ and $\beta$:
\begin{enumerate}
\item $\alpha+\beta = 0$,
\item $\alpha+\beta = \pi/2$.
\end{enumerate}

Next, consider state
$$
{\mathsf T} = \sum_k |c_k|^2 \bigl(|\phi_k\rangle \langle \phi_k|\bigr)
\otimes \bigl(|\phi'_k\rangle \langle \phi'_k|\bigr)\ .
$$
For the projections ${\mathsf P}_k$ and ${\mathsf P}'_k$, all averages in
${\mathsf T}$ equal to those in $\Phi$ and we have again
$$
\rho({\mathsf P}_k,{\mathsf P}'_k,{\mathsf T}) = 1\ .
$$
However, for the observables ${\mathsf P}_{\alpha kl}$ and ${\mathsf
P}'_{\alpha kl}$, we now obtain
$$
\langle {\mathsf P}_{\alpha kl} \rangle_{\mathsf T} = \langle {\mathsf
P}'_{\alpha kl} \rangle_{\mathsf T} = 0\ ,
$$
$$
\Delta_{\mathsf T} {\mathsf P}_{\alpha kl} = \Delta_{\mathsf T} {\mathsf
P}'_{\alpha kl} = \sqrt{|c_k|^2 + |c_l|^2}\ ,
$$
$$
\langle {\mathsf P}_{\alpha kl} {\mathsf P}'_{\beta kl}\rangle_{\mathsf T} =
0\ .
$$
Hence,
$$
\rho({\mathsf P}_{\alpha kl}, {\mathsf P}'_{\beta kl},{\mathsf T}) = 0\ .
$$

Let us summarise: Correlations between ${\mathsf P}_{\alpha kl}$ and ${\mathsf
P}'_{\beta kl}$ determine state $\Phi$ uniquely. The change from $\Phi$ to
${\mathsf T}$ preserves the correlations between ${\mathsf P}_k$ and ${\mathsf
P}'_k$ but erases all correlations between ${\mathsf P}_{\alpha kl}$ and
${\mathsf P}'_{\beta kl}$.

\subsection*{Acknowledgements} The author is indebted to \v{S}tefan
J\'{a}no\v{s} for invaluable help with experimental physics and to Heinrich
Leutwyler and Ji\v{r}\'{\i} Tolar for useful discussions. Thanks go to an
anonymous reviewer for turning attention to the literature on many-body
theory.


\begin{thebibliography}{99}
\bibitem{ludwig1}G. Ludwig, {\it Foundations of Quantum Mechanics I},Springer,
New York, 1983; {\it Foundations of Quantum Mechanics II}, Springer, New York,
1985.
\bibitem{ludwig2}G. Ludwig, {\it An Axiomatic Basis for Quantum Mechanics 1},
Springer, Berlin, 1985; {\it An Axiomatic Basis for Quantum Mechanics 2},
Springer, Berlin, 1987.
\bibitem{zeh}D.~Giulini, E.~Joos, C.~Kiefer, J.~Kupsch, I.-O.~Stamatescu,
H.~D.~Zeh, {\it Decoherence and the Appearance of Classical World in Quantum
Theory}, Springer, Berlin, 1996.
\bibitem{zurek}W.~H.~Zurek, Rev.\ Mod.\ Phys.,{\bf 75} (2003) 715.
\bibitem{hepp}K.~Hepp, Helvetica Phys.~Acta, {\bf 45} (1972) 237.
\bibitem{primas}H. Primas, {\it Chemistry, Quantum Mechanics and
Reductionism.} Springer, Berlin, 1983.
\bibitem{bub}J. Bub, {\it Interpreting the Quantum World}, Cambridge
University Press, Cabridge, UK, 1999.
\bibitem{BLM}P. Busch, P. J. Lahti and P. Mittelstaed, {The Quantum Theory of
Measurement}, Springer, Heidelberg, 1996.
\bibitem{d'Espagnat}B. d'Espagnat, {\em Veiled Reality}, Addison-Wesley,
Reading, 1995.
\bibitem{PHJT}P. H\'{a}j\'{\i}\v{c}ek and J. Tolar, Found.\ Phys.\ {\bf 39}
(2009) 411.
\bibitem{hajicek}P. H\'{a}j\'{\i}\v{c}ek, Foud.\ Phys.\ {\bf 39} (2009) 1072.
\bibitem{belt}E. G. Beltrametti, G. Cassinelli and P. J. Lahti, J. Math.\
Phys.\ {\bf 31} (1990) 91.
\bibitem{peres}A.~Peres, {\it Quantum Theory: Concepts and Methods}, Kluwer,
Dordrecht, 1995.
\bibitem{JvN}J.~von~Neumann, {\it Mathematical Foundation of Quantum
Mechanics}, Princeton University Press, Princeton NJ, 1983.
\bibitem{bohm}D. Bohm, {\it Quantum Theory}, Prentice-Hall, Englewood Cliffs,
1951.
\bibitem{leo} W. R. Leo, {\it Techniques for Nuclear and Particle Physics
Experiments}, Springer, Berlin, 1987.
\bibitem{stefan}D. Twerenbold, Rep.\ Progr.\ Phys. {\bf 59} (1996) 239.
\bibitem{Weinberg}S.~Weinberg, {\em The Quantum Theory of Fields} Vol.\ I,
P. 177.  Cambridge University Press, Cambridge 1995.
\bibitem{haag}R. Haag, {\it Local Quantum Physics. Fields, Particles,
Algebras}. Springer, Berlin, 1992.
\bibitem{KP}B. D. Keister and W. N. Polyzou, in {\it Advances in Nuclear
Physics}, ed.\ J. W. Negele and E. Vogt, Plenum, New York 2002. Vol 20.
\bibitem{coester}F. Coester, Int.\ J. Modern Phys.\ {\bf 17} (2003) 5328.
\bibitem{mott}N. F. Mott, Proc.\ Roy.\ Soc.\ London Series A {\bf 126} (1929)
79.
\bibitem{DG}C. Davisson and L. Germer, Phys.\ Rev.\ {\bf 30} (1927).
\bibitem{MW}L. Mandel and E. Wolf, {\it Optical Coherence and Quantum Optics},
Cambridge University Press, Cambridge, 1995.
\end{thebibliography}
\end{document}